\documentstyle[twoside,fleqn,epsfig,float,espcrc2]{article}
\newcommand{\Dirac}{\mbox{$\not\!\!D$}}

\title{Equivalent of a Thouless energy in lattice QCD Dirac spectra}

\author{M.E.~Berbenni\address{Fachbereich Physik, Universit\"at
    Kaiserslautern, D-67663 Kaiserslautern, Germany}\thanks{Poster 
  presented by M.E.~Berbenni},
  T.~Guhr\address{Max Planck Institut f\"ur Kernphysik,
  D-69029 Heidelberg, Germany},
  J.-Z.~Ma\address{Hong Kong Baptist University, Hong Kong},
  S.~Meyer$^{\rm a}$,
  T.~Wilke$^{\rm b}$}

\begin{document}
\begin{abstract}
Random matrix theory (RMT) is a powerful statistical tool to model 
spectral fluctuations. 
In addition, RMT provides efficient means to separate different scales 
in spectra. 
Recently RMT has found application in quantum chromodynamics 
(QCD). 
In mesoscopic physics, the Thouless energy sets the universal scale for 
which RMT applies. 
We try to identify the equivalent of a Thouless energy in complete 
spectra of the QCD Dirac operator with staggered fermions and $SU_c(2)$ 
lattice gauge fields. 
Comparing lattice data with RMT predictions we find deviations which 
allow us to give an estimate for this scale. 
\end{abstract}

\maketitle

\begin{figure}[t]
\vspace*{3mm}
  \epsfig{figure=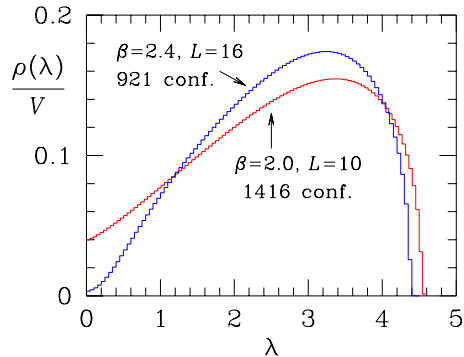,width=60mm}
\vspace*{-6mm}
  \caption{Spectral density of the lattice Dirac operator for 
  $\lambda > 0$ for two different lattice volumes and gauge coupling 
  values.}
  \label{fig1}
\vspace*{-8mm}
\end{figure}

In recent years, RMT has been successfully introduced into the study 
of certain aspects of quantum chromodynamics (QCD). The interest 
focuses on the spectral properties of the Euclidean Dirac operator.
For the massless Dirac operator $\Dirac[U]$ with staggered fermions and 
gauge fields $U \in SU_c(2)$ we solve numerically for each configuration
the eigenvalue equation
\begin{equation}
  \label{eq1} 
  i\Dirac[U] \psi_k=\lambda_k[U] \psi_k. 
\end{equation}
The distribution of the gauge fields is given by the Euclidean 
partition function.  
Examples of the spectra are shown in Fig.~\ref{fig1}, where the
average level densities for $16^4$ and $10^4$ lattices are shown.
It should be pointed out that we have $V/2=32768$ resp. $V/2=5000$ 
distinct positive eigenvalues of each configuration, so that there 
are millions of eigenvalues at our disposal.

As the gauge fields vary over the ensemble of configurations, the 
eigenvalues fluctuate about their mean values. Chiral random matrix 
theory models the fluctuations of the eigenvalues in the microscopic 
limit, i.e. near $\lambda=0$ \cite{1} as well as in the bulk of the 
spectrum \cite{2}. Our main question is to what scales RMT does 
apply in QCD. In disordered systems the Thouless energy $E_c$ 
determines the scale in which fluctuations are predicted by RMT. 
Beyond this scale deviations occur. In QCD an equivalent of the 
Thouless energy is $\lambda_{\rm RMT}$ \cite{3}. In the microscopic 
region it scales as 
\begin{equation}
\lambda_{\rm RMT}/D \propto \sqrt V,
\end{equation} 
$V$ is the lattice volume, $D$ is the mean level spacing. 
As argued in \cite{osborn98} a corresponding effect should also be 
seen in the bulk of the spectrum.\\
\begin{figure*}
\vspace*{-10mm}
  \centerline{\epsfig{figure=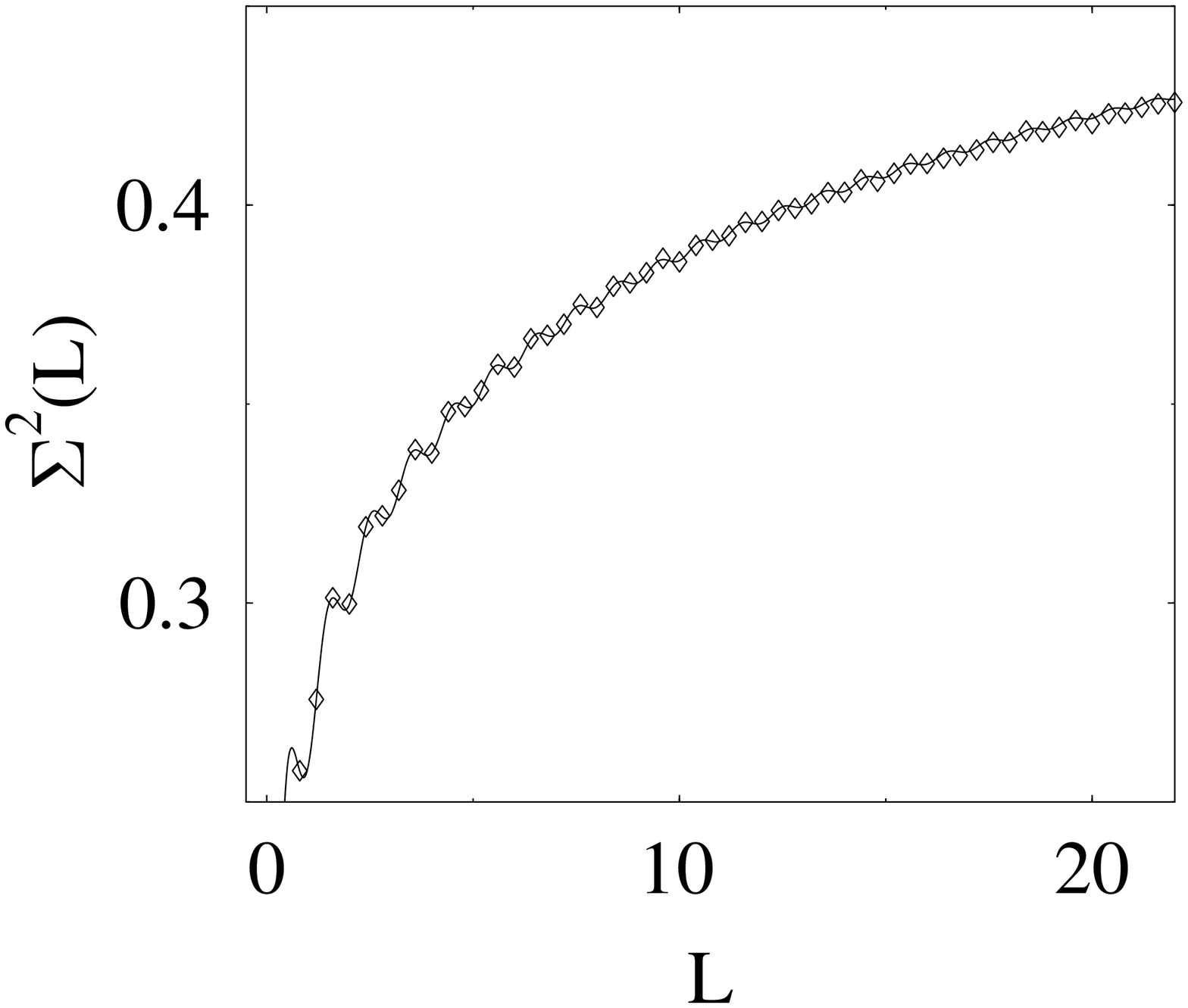,width=60mm}
  \hspace*{5mm}\epsfig{figure=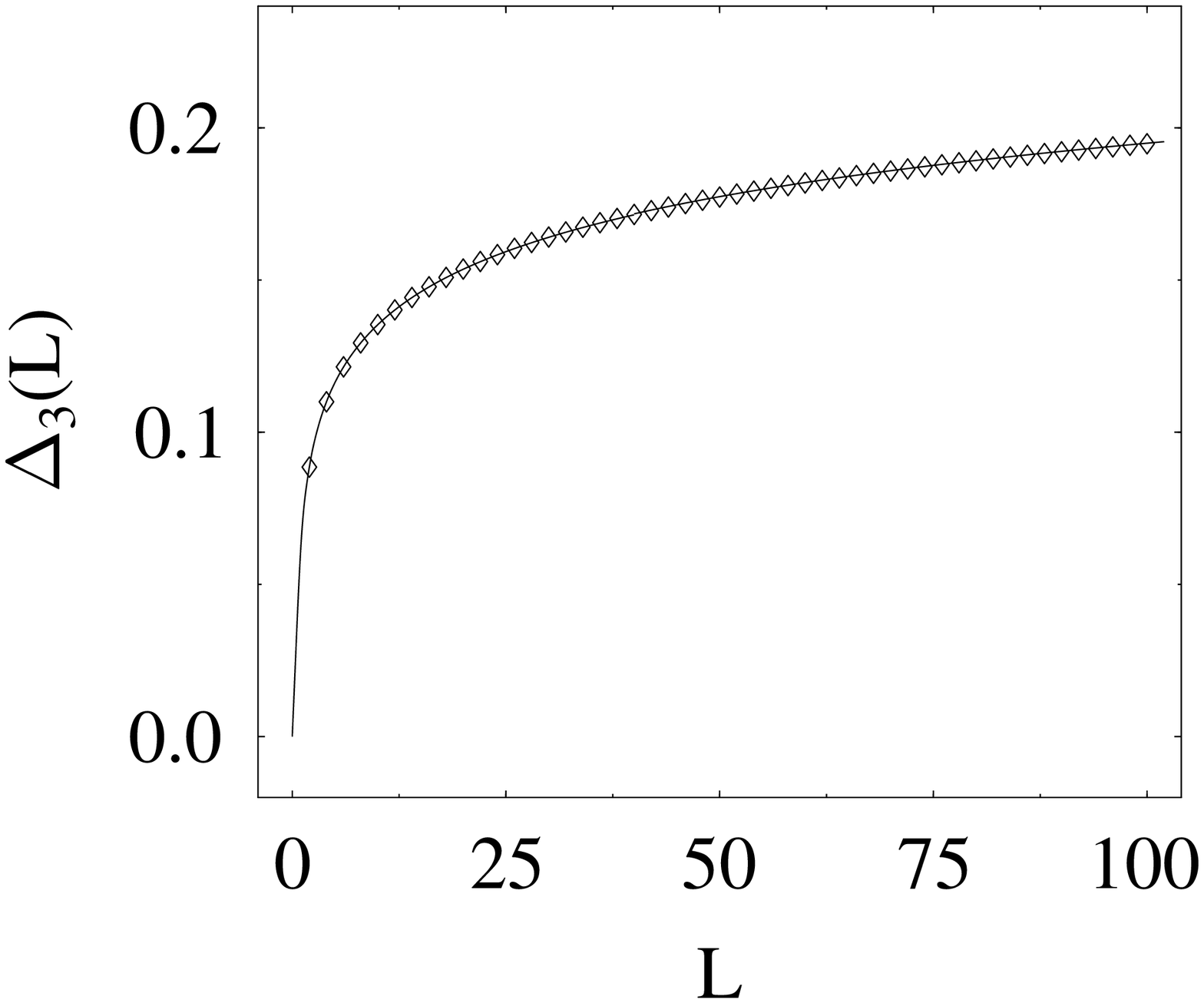,width=60mm}}
  \vspace*{-10mm}
  \caption{Number variance (left) and spectral rigidity (right)
  on a $16^4$ lattice.
  The solid lines are RMT predictions, the dots numerical data.
  On this scale the data points do not depend on unfolding.}
  \label{fig2}
\end{figure*}
The staircase function $N(\lambda)$ gives the number 
of levels with energy $\le \lambda$. In many cases it can be 
separated into
\begin{equation}
  \label{eq2} 
 N(\lambda)=N_{\rm ave}(\lambda)+N_{\rm fluc}(\lambda).
\end{equation}
$N_{\rm ave}(\lambda)$ is determined by gross features of the system.
$N_{\rm fluc}(\lambda)$ contains the correlations to be analyzed.
RMT makes predictions for the fluctuations on the scale of the 
mean level spacing. The influence of the overall level density 
must be removed by numerically unfolding the spectra through 
the mapping $\lambda_i \rightarrow x_i = N_{\rm ave}(\lambda_i)$.
For the new sequence we then have $\hat{N}_{\rm ave}(x)=x$
i.e. the mean level spacing is unity everywere 
$1/\rho_{\rm ave}(x)=1$ where 
$\rho_{\rm ave}(x)=d\hat{N}_{\rm ave}(x)/dx$.
The extraction of $N_{\rm ave}(\lambda)$ is highly non-trivial, 
because little is known analytically about the level density of 
QCD spectra. 

However, there are several phenomenological unfolding procedures, 
e.g. ensemble unfolding, where one divides the energy range in $m$ 
bins of width $\Delta\lambda$ and averages the density 
$\rho(\lambda,\lambda+\Delta\lambda)$ for each bin over all 
configurations. Then the staircase function 
$N_{\rm ave}(\lambda)=\sum_{i=1}^m\rho(\lambda_i,\lambda_i+\Delta\lambda)
\Delta\lambda$, with $\lambda_m=\lambda$ is calculated. Furthermore 
there is configuration unfolding, where $N(\lambda)$ is fitted for 
each configuration to a polynomial of degree $n$. 
Strong coupling expansions for $SU_c(2)$ with staggered fermions 
\cite{kogut83} and $1/N_c$ expansion of the QCD level density 
\cite{smilga93} motivate this ansatz. For technical 
details and further unfolding procedures see \cite{4}.
Whatever approach one uses, the mean number of rescaled levels in 
an interval of length $L$ in units of the mean level spacing should 
equal $L$. This assures that the unfolded spectrum has mean level 
density unity.

We compared RMT predictions for two-point correlators with 
lattice data for two quantities. First, the level number 
variance, which measures the deviation of the number of 
eigenvalues $n_\alpha(L)$ in an interval $[\alpha,\alpha+L]$ 
from the expected mean number $L$ 
\begin{equation}
 \label{eq7}
 \Sigma^2(L)=\overline{<(L-n_{\alpha}(L))^2>}. 
\end{equation}
$< \cdots >$ is the spectral average, $\overline{( \cdots)}$ 
the ensemble average. 
Thus, an interval of length $L$ contains on average 
$L \pm \sqrt{\Sigma^2(L)}$ levels. For uncorrelated Poisson 
spectra $\Sigma^2(L)=L$. RMT predicts stronger correlations: 
$\Sigma^2(L) \sim \log L$. 
The second two-point correlator we considered is the spectral 
rigidity, defined as the least square deviation of $N(\lambda)$ 
from the straight line
\begin{equation}
 \label{eq8}
 \Delta_3(L)=\left<\frac{1}{L}{\rm
    min}_{A,B}\!\!\int_{\alpha}^{\alpha+L}\!\!\!\!\!\!\!\!\!\!
    d\xi(N(\xi)-A\xi-B)^2\right>\!\!.
\end{equation} 
For this quantity RMT predicts $\Delta_3(L) \sim \log L$. In Fig.~\ref{fig2} 
the RMT results for these statistical measures are compared with 
lattice data. The wealth of data allows us to analyze higher order 
correlations. Again we see good agreement (see Figs. 11 and 12 
in \cite{4}).
\begin{figure*}
\vspace*{-8mm}
   \centerline{\epsfig{figure=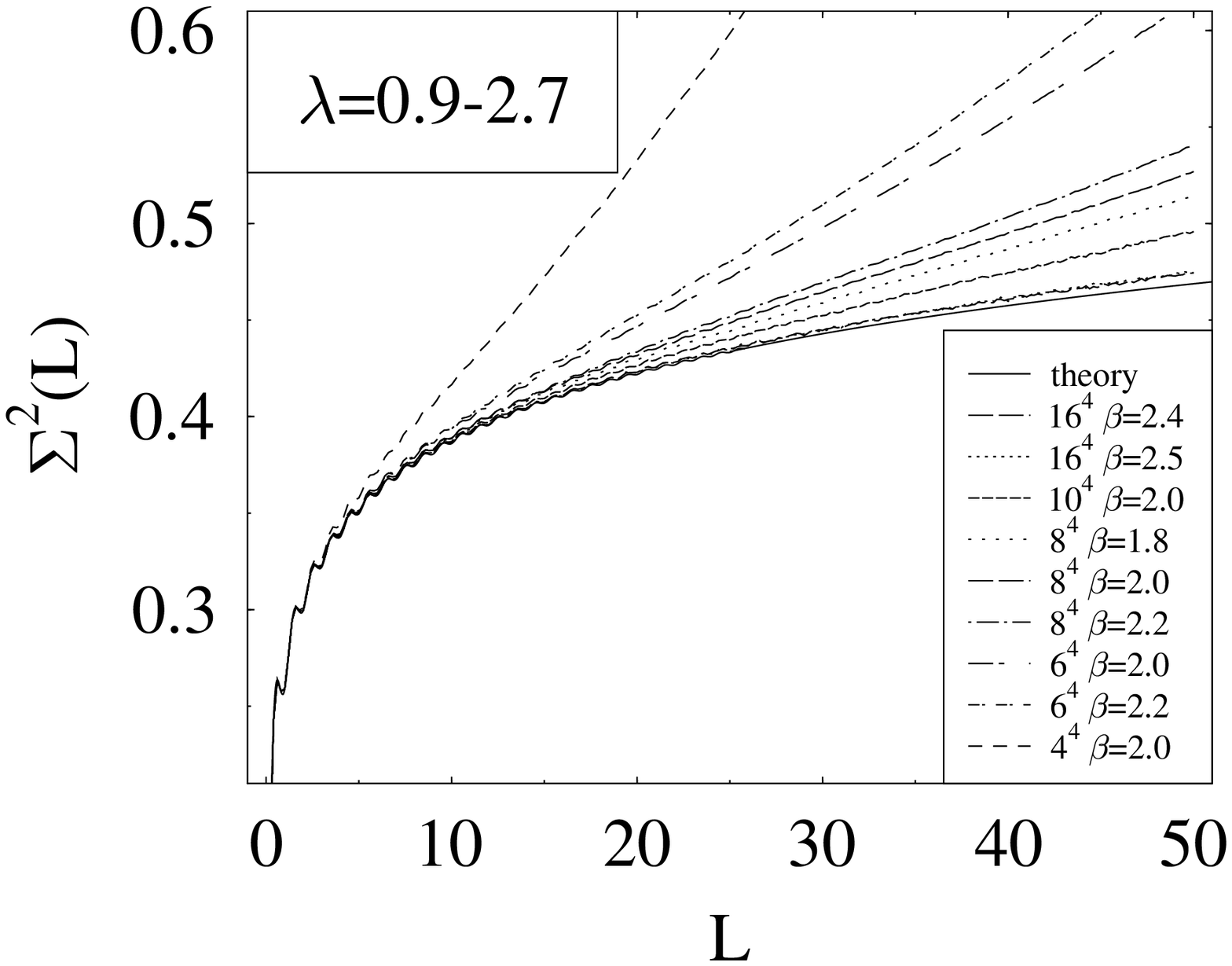,width=70mm,angle=0}
   \hspace*{5mm}
   \epsfig{figure=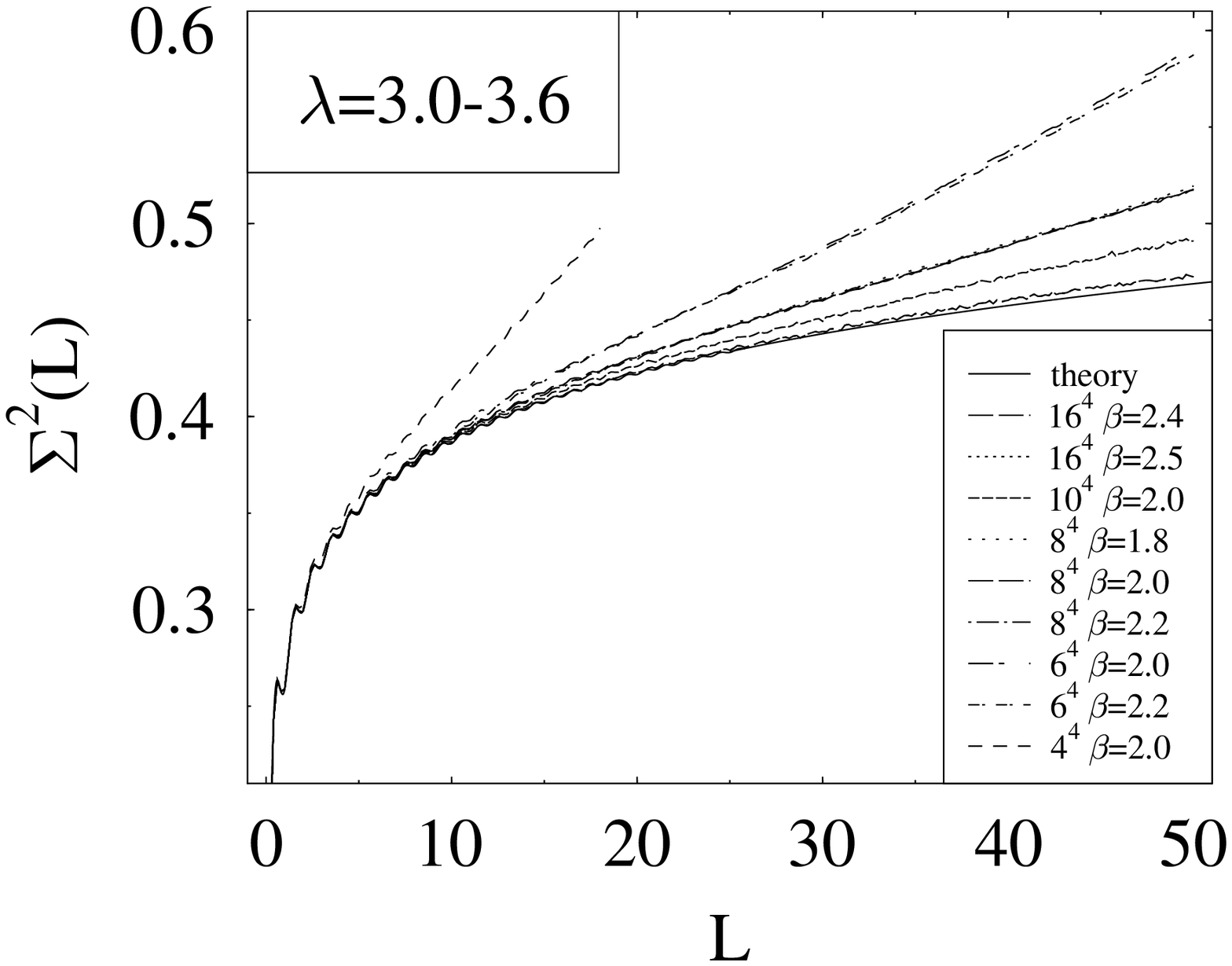,width=70mm,angle=0}} 
\vspace*{-8mm}
  \caption{Deviations from RMT predictions for different lattices
         sizes $V$ and gauge couplings $\beta$. Shown are different
         regions of the spectrum as indicated in the upper left
         part of the plots.}         
  \label{fig3}
\vspace*{-3mm}
\end{figure*}

With ensemble unfolding of the data we obtain for $\Sigma^2(L)$ the
curves plotted in Fig.~\ref{fig3}. Independently of the spectral 
region considered \cite{4} and of $\beta$ we find that the point 
where the deviation sets in, scales as
\begin{equation}
 \label{eq9}
 \lambda_{\rm RMT}/D \simeq 0.3 \sqrt V. 
\end{equation}
This should be compared with the result obtained in \cite{3} 
for the microscopic region
\begin{equation}
 \label{eq10}
 \lambda_{\rm RMT}/D \simeq 0.3 \dots 0.7 \sqrt V. 
\end{equation}

\begin{figure}[ht]
  \epsfig{figure=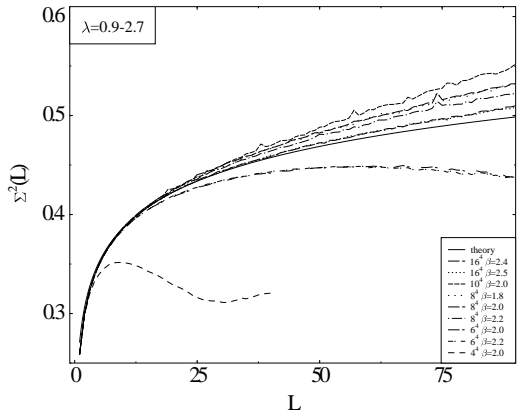,width=85mm}
\vspace*{-15mm}
  \caption{Deviations from RMT predictions with configuration
   unfolding of the data for different lattice sizes and $\beta$ 
   values.}
  \label{fig4}
\vspace*{-8mm}
\end{figure}

With polynomial unfolding the scaling law (\ref{eq9}) vanishes. 
The deviation point appears to be the same for different lattice 
sizes (Fig.~\ref{fig4}).

In order to find out, 
if these deviations are due to a Thouless energy, we performed 
a Fourier analysis of the oscillations of the staircase function 
\cite{4}. From it we concluded that the deviations of the data 
obtained with polynomial unfolding are due to a non-polynomial-like 
part in the average level density and not to an equivalent of the 
Thouless energy.

In conclusion, analyzing some of the statistical properties of 
complete eigenvalue 
spectra of the Dirac operator for staggered fermions and $SU_c(2)$ 
gauge fields for various couplings and lattice volumes, we find
the scaling behavior of the equivalent of the Thouless energy.
Using ensemble unfolding, we have $\lambda_{\rm RMT}/D= C\sqrt V$. 
The constant is approximately $C \approx 0.3$ which is compatible 
with the result obtained in \cite{3} for the microscopic region of 
the spectrum, where the scaling (\ref{eq10}) was found. 
By unfolding each configuration separately, we do not see any scaling 
of this type. Hence the Thouless energy is due to fluctuations 
in the ensemble.   

\end{document}